\documentclass[conference]{IEEEtran}
\IEEEoverridecommandlockouts
\usepackage[ruled]{algorithm2e}
\usepackage{cite}
\usepackage{algorithmic}
\usepackage{caption}
\usepackage{amsmath,amssymb,amsfonts}
\usepackage{graphicx}
\usepackage{textcomp}
\usepackage{xcolor}
\usepackage{float}
\def\BibTeX{{\rm B\kern-.05em{\sc i\kern-.025em b}\kern-.08em
    T\kern-.1667em\lower.7ex\hbox{E}\kern-.125emX}}
\begin{document}

\title{ Comparative study of performance of parallel Alpha Beta Pruning for different architectures 
}

\author{\IEEEauthorblockN{Shubhendra Pal Singhal}
\IEEEauthorblockA{\textit{Department of Computer Science and Engineering} \\
\textit{National Institute of Technology, Tiruchirappalli}\\
shubhendrapalsinghal@gmail.com}
\and
\IEEEauthorblockN{M. Sridevi}
\IEEEauthorblockA{\textit{Department of Computer Science and Engineering} \\
\textit{National Institute of Technology, Tiruchirappalli}\\
msridevi@nitt.edu}
}

\maketitle

\begin{abstract}
Optimization of searching the best possible action depending on various states like state of environment, system goal etc. has been a major area of study in computer systems. In any search algorithm, searching best possible solution from the pool of every possibility known can lead to the construction of the whole state search space popularly called as minimax algorithm. This may lead to a impractical time complexities which may not be suitable for real time searching operations. One of the practical solution for the reduction in computational time is Alpha Beta pruning. Instead of searching for the whole state space, we prune the unnecessary branches, which helps reduce the time by significant amount. This paper focuses on the various possible implementations of the Alpha Beta pruning algorithms and gives an insight of what algorithm can be used for parallelism. Various studies have been conducted on how to make Alpha Beta pruning faster. Parallelizing Alpha Beta pruning for the GPUs specific architectures like mesh(CUDA) etc. or shared memory model(OpenMP) helps in the reduction of the computational time. This paper studies the comparison between sequential and different parallel forms of Alpha Beta pruning and their respective efficiency for the chess game as an application.   
\end{abstract}

\begin{IEEEkeywords}
Parallel algorithms, Minimax, Alpha Beta pruning, CUDA, OpenMP, Mesh architecture, Shared memory model
\end{IEEEkeywords}

\section{Introduction}
Playing a game strategically requires an individual to foresee all kinds of winning possibilities. The grading policy applied to game tree is generally +1 for winning and -1 for losing which ultimately helps the agent decide the next move. This may require the construction of whole state space implying that every possibility or action needs to be considered and whatever suits the best or takes the agent close to its goal, should be opted. Now, this is the general brute force method which might work practically for simpler and smaller games like Tic-Tac-toe, but the complex games like checkers or chess on 8*8 board, has a gigantic state space and searching using the brute force method in such a huge space is impractical. This gives us the motivation to study a better and feasible method called Alpha Beta pruning. The cases which turns out to be futile at the start of the search are rejected, thus reducing the state space for every move by significant amount. Furthermore, the efficiency of the Alpha Beta pruning can be further improved, to suit feasibility of running AI applications\cite{eChronos13} consisting of accrued state space. Parallelism turns out to be one of the factors which can be used to improve the performance. This paper compares the performances and speedup obtained from various implementations of the parallel forms of Alpha Beta pruning on different architectures.\newline 
The earlier the branches are pruned, the better is the efficiency of Alpha Beta pruning. Different architectures prune the branches at different time stamps, thus differing in the computation time.

\subsection{Minimax Algorithm}
Brute force search in the state space is the minimax algorithm. Minimax is a decision-making algorithm \cite{eChronos}. The main aim of the algorithm is to find the next best move as shown in Fig1 for the game Tic-Tac-Toe\cite{eChronos6} (application of minimax) .\newline
\begin{figure}
    \includegraphics[scale = 0.4]{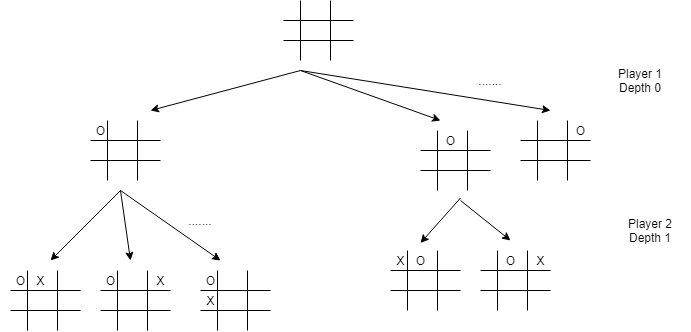}
    \caption{Game tree for Tic-Tac-toe}
\end{figure}{} \newline
In the applications of Minimax algorithm (involving two players), first player is maximizer, and the second player is the minimizer. The evaluation score is assigned to the game board, where the maximizer and minimizer aims to get the highest score, and the lowest score possible respectively \cite{eChronos}. The implementation of the same in shown in the following algorithm 1\cite{eChronos8}.\newline \newline
\begin{algorithm}
\caption{Minimax Algorithm}
function minimax(node,depth, maximizingPlayer)\\
 \If{$depth = 0$ or node is a leaf node}{
            return heuristic value of node
            }
\If{$maximizingPlayer$}{
    value = -$\infty$ \\
    \While{every child of node}{
        value = max(value, minimax(child, depth-1, FALSE)
    }
    return value
 }
\Else{
    value = +$\infty$ \\
    \While{every child of node}{
        value = min(value, minimax(child, depth-1, TRUE)
    }
    return value
 }
\end{algorithm}{}
If every child branch is allocated to one processor and run the same algorithm in parallel for every child branch, then the parallel minimax is formed. The parent child will collect the answer from all the child branches and further propagate the optimal one to its ancestors till it finally converges to one parent i.e. root of the tree.

\subsection{Alpha Beta Pruning}
Alpha–Beta pruning is a search algorithm that seeks to decrease the number of nodes that are evaluated by the minimax algorithm in its search tree. It is an adversarial search algorithm which stops evaluating a move when at least one possibility has been found that proves the move to be worse than a previously examined one. Such moves need not be considered further. In this way, this leads to pruning of the branch as this will never affect the final decision that agent has to make to achieve the goal. 
\subsubsection{Sequential Alpha Beta Pruning}
The Alpha-Beta pruning algorithm emerged as an
improvement over the usual Minimax algorithm. Both of these algorithms find out the best
available move to the player and will return the exact
utility regarded with that move. The execution time of Alpha-Beta
pruning algorithm is comparatively faster than
Minimax algorithm as they cut down(prune) the branches, thereby
reducing the exploration space. The reason being, that the
values calculated from these branches would
not affect the final result. Since, time is not spent on
exploring the other branches, these algorithms
effectively abates the execution time.
The implementation of the same in shown in the following algorithm 2, with an illustration shown in Fig 2.

\subsubsection{Parallel Alpha Beta Pruning}
Parallelizing Alpha Beta pruning can further reduce the execution time and improves the performance. The insight to the solution is to parallelize the searching of branches which are to be pruned i.e. the
evaluation of the game tree is parallelized.
Essentially, each branch of the game tree can be
evaluated in parallel. Thus, the Alpha and Beta values
are propagated at once to each of the node at the first
level and the minimizing and corresponding
maximizing moves are evaluated simultaneously for
all the branches. The final result of all the
branches is propagated and is handled by the parent and the best
move is computed 
\begin{algorithm}
\caption{Sequential Alpha Beta Pruning Algorithm}
function ALPHA-BETA(node, depth, $\alpha$, $\beta$, maximizingPlayer) \\
\If{$depth = 0$ or node is a leaf node}{
            return heuristic value of node
            }
\If{$maximizingPlayer$}{
    value = -$\infty$ \\
    \While{every child of node}{
        value = max(value, ALPHA-BETA(child, depth-1,$\alpha$, $\beta$, FALSE)\\
        $\alpha$ = max($\alpha$, value)\\
        \If{$\beta \leq \alpha$ }{break}
    }
 return value
 }
\Else{
    value = +$\infty$ \\
    \While{every child of node}{
        value = min(value, ALPHA-BETA(child, depth-1,$\alpha$, $\beta$, TRUE)\\
        $\beta$ = min($\beta$, value)\\
        \If{$\beta \leq \alpha$ }{break}
    }
    return value
 }
\end{algorithm}{} 
\begin{figure}
    \includegraphics[scale =  0.35]{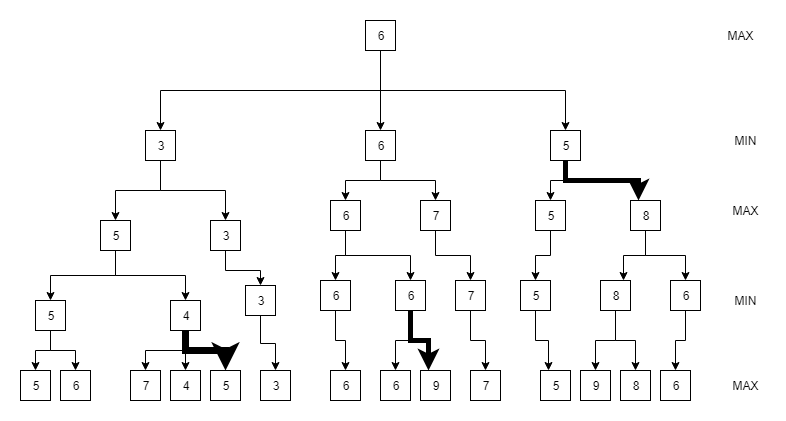}
    \caption{Illustrating the sequential Alpha-Beta pruning\\ \\ \tiny The highlighted arrows represents the cut that was being unnecessarily being explored in Minimax. This Alpha-Beta cut ensures that the branch is not explored further, because it is a mere waste of computation time. The highlighted arrows represents those branches which will not be further explored by Alpha-Beta Pruning algorithm, thus making it more efficient as it is evident from the last cut which saves up massive amount of time.}
\end{figure}{}
with respect to the Alpha and Beta
values of each branch.\newline
The partitioning of the tree for parallel computation is
done on a per-child basis\cite{eChronos3}. Each child of the
tree evaluates the minimum and maximum moves
together which is independent of the other branches(children of same parent) and thus computation of every branch is relatively faster\cite{eChronos3}. After this computation is
done, the parent decides upon the final result. The implementation of the same is given in algorithm 3.
\newline \newline
\begin{algorithm}
\caption{Parallel Alpha Beta Pruning Algorithm}
function Parallel-ALPHA-BETA(node, depth, $\alpha$, $\beta$, maximizingPlayer) \\
 \If{$depth = 0$ or node is a leaf node}{
            return heuristic value of node\\
            }
\If{$maximizingPlayer$}{
    value = -$\infty$ \\
    \While{every child of node in parallel}{
        value = max(value, Parallel-ALPHA-BETA(child, depth-1,$\alpha$, $\beta$, FALSE)\\
        $\alpha$ = max($\alpha$, value)\\
        \If{$\beta \leq \alpha$ }{break}
    }
 return value
 }
 \Else{
    value = +$\infty$ \\
    \While{every child of node in parallel}{
        value = min(value, Parallel-ALPHA-BETA(child, depth-1,$\alpha$, $\beta$, TRUE)\\
        $\beta$ = min($\beta$, value)\\
        \If{$\beta \leq \alpha$ }{break}
    }
    return value
 }
\end{algorithm}{} 
\newline
Over past few decades, proposal of implementing parallel Alpha Beta pruning on different architectures have been proposed\cite{eChronos1,eChronos3,eChronos4,eChronos9}. One amongst them is the traditional algorithm which uses the prioritizing scheme for improving further efficiency. Principal-Variation algorithm\cite{eChronos7} is one such attempts, which states the prioritizing scheme as the rule of
searching the first branch at a PV node before the search of remaining branches begins. Since the implementation of PV Split algorithm on the existing architecture, there has been various possible optimization techniques\cite{eChronos3} that have been suggested based on using that prioritizing scheme carefully and efficiently. All these schemes have been proposed and are being used in different environments like beam search, reordering being used in chess AI inbuilt Windows game\cite{eChronos5}. But the real question that is never answered is that which method is the optimal under the specified constraints? So, this paper attempts to implement different parallel models for Alpha Beta pruning and also compare the results to find the optimal solution for the problem.

\section{Experimental setup and implementation}
There are various platforms available which allows the parallel implementation of an algorithm. Shared memory model and architectures like mesh have been used for the comparison. All the algorithms have been implemented for application of two player game chess for different board sizes\cite{eChronos6}.
\subsection{Prerequisites for the implementation}
For implementing the algorithm, the following setup is required.\newline
\begin{itemize}
    \item[1] A computing machine with 8GB RAM and Linux environment.
    \item[2] Installation of OpenMP.
    \item[3] Installation of NVIDIA CUDA Toolkit for Linux environment.
\end{itemize}

\subsection{Algorithm used for implementation of Parallel Alpha Beta pruning}
The way, parallelism is introduced in the Alpha Beta algorithm is by developing the concurrent processing program
of multiple child nodes at every level of the minimax
tree\cite{eChronos19}. Amongst all the moves, if the best move can be evaluated first, then the
rest of the moves can be straightaway ignored. Unfortunately, the a priori quality of estimation might super-cede the time taken by sequential Alpha Beta pruning.
The basis for our parallel Alpha-Beta implementation is the root splitting algorithm\cite{eChronos7}.
The main idea of this algorithm is to ensure that each node except for the root, is allocated to only one processor. To keep the effect of the Alpha Beta pruning, we split the children nodes of the root into clusters, each cluster being served by only one processor. Each child is then processed as the root of a sub search tree that will be visited in a sequential fashion. When a child finishes computing it returns its result to the root node, and the root node decides the best result. The mutual exclusion and the critical section constraints are taken into consideration while implementation.\newline
Along with root splitting, one another optimization is included. Reordering the children of a given node in a way to start exploring the most promising branches, thus pruning of the unwanted branches earlier making the algorithm more faster.\newline
Further, this algorithm is compared to one of the optimization i.e. beam search\cite{eChronos5} which uses breadth-first search to construct a search tree. At every level of the search tree, it produces all the successors of the states at the current level and then sorts them in increasing order of their heuristic costs.\newline
The algorithm only explores a subset of those child nodes, by ignoring the least promising ones (according to the estimation function), and thus making the search faster.

\subsection{Mesh Architecture - CUDA}
The mesh architecture\cite{eChronos20} of size n is a machine with n simple processors arranged in a square lattice. Processor P(i,j) represents the processor in row i and column j in mesh as shown in Fig3.
\begin{figure}
    \includegraphics[scale = 0.5]{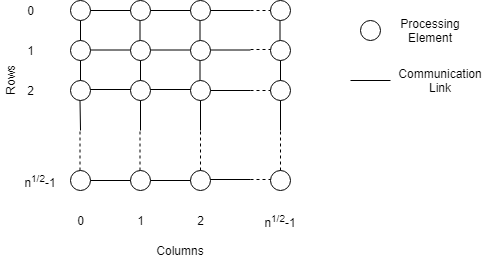}
    \caption{Mesh Architecture of size n}
\end{figure}{} \newline
CUDA is a parallel computing framework created by Nvidia. The CUDA platform is a software layer that gives direct access to the virtual instruction set and parallel computational elements, for the execution of compute kernels. It is used for the implementation of parallel Alpha Beta pruning on mesh architecture.\newline
The parallel architecture of the Nvidia GPUs is made up of a multiple sets of pipelined multiprocessors. The
parallel computation on the GPU is performed
in a similar fashion, as a set of concurrently executing thread blocks, which are organized into a 1D or 2D grid\cite{eChronos4}.
\subsubsection{Implementation}
In CUDA, the parallel processing of child nodes is done using multiple thread blocks(mesh architecture).

The parts of the root split algorithm executed in parallel by all of the processors are node evaluation and optimal move generation.\cite{eChronos1,eChronos4}.

\subsection{Shared Memory Model - OpenMP}
Shared memory system has single (shared) space accessible by multiple processors but each process has its own address space which is not accessible by other processes.\cite{eChronos21}\newline 
OpenMP (Open Multi-Processing) is an application programming interface (API) that supports multi-platform shared memory multiprocessing programming.\newline
It is a framework with the model of parallel programming which runs on a computer cluster. OpenMP is used for parallelism within a multi-core node. It is used for the implementation of shared memory model of parallel Alpha Beta pruning. 
\subsubsection{Implementation}
Every part of root split algorithm that is intended to keep running
in parallel is assigned a thread from the pool of forerun created free threads. Every thread assigned is in the "running" state which propagates the result to the parent thread which ultimately decides that which branch is optimal.\newline
As a matter of course, every thread executes
the parallelized segment of code freely using the pragma directive in omp library. The thread assigned to the pruned branch is assigned "free" state and is allocated back to the pool of forerun created free threads. They can be further be utilised by other segments, thus optimizing the memory usage and execution time.\cite{eChronos3} 

\section{Experimental Results}
The first subsection of results computes the relative speedup between the two optimizations used in parallel alpha beta pruning. The optimization which would turn out to be better, will be used as the algorithm for the next section where the architectures used for the implementation have been compared.
\subsection{Reordering vs Beam Search}
\begin{figure}
\begin{center}
\includegraphics[scale = 0.5]{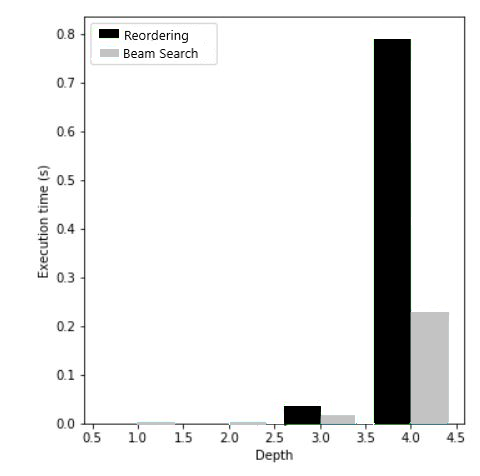}
\end{center}{}
\caption{Execution time of Reordering vs Beam Search}
\end{figure}{}
From Fig4, it can be deduced that average speedup of 2.45 can be obtained relative to the reordering technique by using the beam search optimization.
There is a direct implication between the number of visited nodes and the execution time. If less number of nodes are visited, then it will lead to better performance. Therefore, beam search optimizes the number of visited number of nodes when compared to the Reordering optimization as shown in the Fig5.\newline
Thus, we conclude that beam search is the optimal implementation where the speedup is highest and the number of visited nodes is also less, inturn improving performance.
\begin{figure}
\begin{center}
\includegraphics[scale = 0.5]{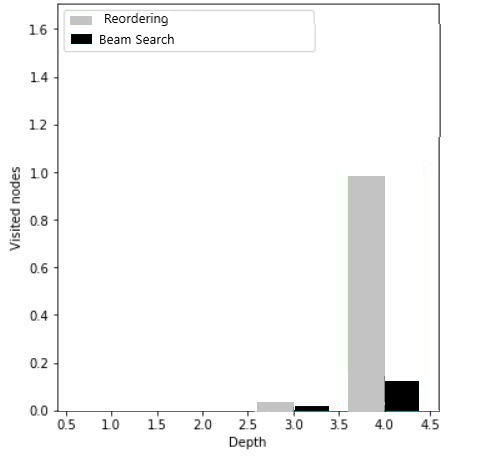}
\end{center}{}
\caption{Number of visited nodes in Beam Search vs Reordering}
\end{figure}{}

\subsection{Mesh Architecture vs Shared Memory Model}
\begin{figure}
\begin{center}
\includegraphics[scale = 0.3]{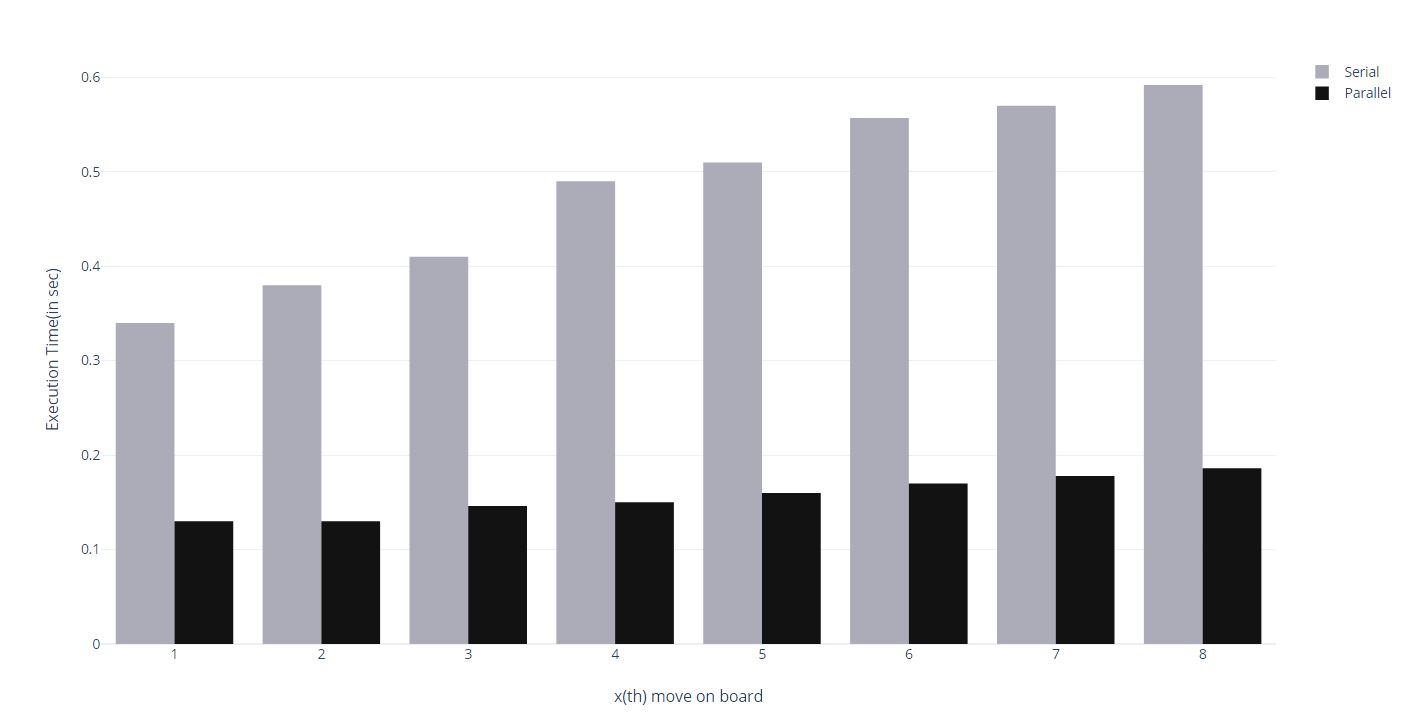}
\end{center}{}
\caption{Parallel (OpenMP) vs sequential Alpha Beta pruning}
\end{figure}{}
Beam Search with root splitting algorithm is used for implementation on both architectures as the previous section concluded that beam search is the best optimization amongst all.
The average speedup achieved by using parallel
execution in OpenMP\cite{eChronos3} i.e. shared memory model is 3.23 as illustrated in Fig6.
\begin{figure}
\begin{center}
\includegraphics[scale = 0.7]{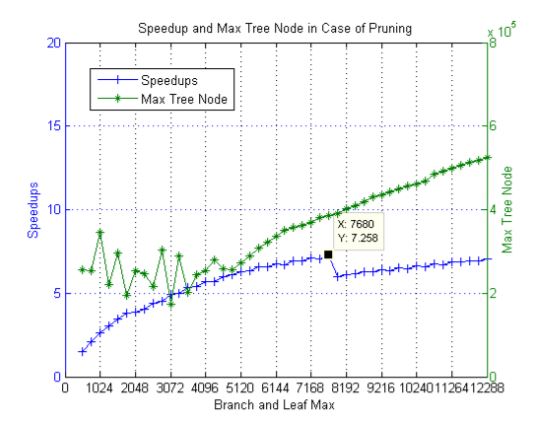}
\end{center}{}
\caption{Parallel (CUDA) vs sequential Alpha Beta pruning}
\end{figure}{}
The average speedup achieved by using parallel
execution in CUDA\cite{eChronos9} (using 4 thread model) is 7.26 as illustrated in Fig7. If we compare the implementations of parallel Alpha Beta pruning between OpenMP and CUDA, CUDA proves to be a better architecture due to possible reasons of efficient multi-processor management and optimized architecture.\newline

\section{Conclusion}
The first section of the paper discussed the sequential forms of minimax and Alpha Beta pruning and proved that Alpha Beta pruning can improve performance. Next section of the paper walks through the implementation of parallel Alpha Beta pruning.  
The algorithm used for the implementation is root split, but this algorithm has been further optimized by using Reordering and beam search. But, different optimization lead to different execution times. Beam Search leads to maximum increase in performance as compared to others. Thus, authors have used beam search for implementation of parallel of Alpha Beta pruning for mesh and shared memory model. 

Not every implementation in parallel scenario of Alpha Beta pruning leads to similar speedup. The speedup for Alpha Beta pruning using mesh architecture(CUDA) proves to be almost 2x faster than using OpenMP. So it is concluded that the implementation of Alpha Beta pruning using beam search in CUDA(using 4 thread model) is better than any other possible combination of implementation like Reordering+OpenMP.\newline

\bibliographystyle{IEEEtran}
\bibliography{IEEEexample.bib}
\end{document}